\documentclass[aps,prd,reprint,nofootinbib,showpacs,groupedaddress]{revtex4-2}

\usepackage{amsmath,amssymb,bm}
\usepackage{graphicx}
\usepackage{booktabs}
\usepackage{microtype}
\usepackage[hidelinks]{hyperref}

\newcommand{\Oa}{\mathcal O_a}
\newcommand{\Ob}{\mathcal O_b}
\newcommand{\Oc}{\mathcal O_c}

\newcommand{\bne}{b_{ne}}
\newcommand{\rn}{\langle r_n^2\rangle}
\newcommand{\lrpartial}{\overset{\leftrightarrow}{\partial}}

\begin{document}



\title{Constraints on Neutron-Electron Quantum Dark Forces}


\author{V. Kurmangaliyeva}
\author{B. Massak}
\author{O. Agyl-Mussapar}
\author{S. Amangeldinova}

\affiliation{al-Farabi Kazakh National University, Almaty, Kazakhstan}

\author{W. M. Snow}
\affiliation{Indiana University/CEEM, 2401 Milo B. Sampson Lane, Bloomington, Indiana 47408, USA}

\date{\today}

\begin{abstract}
Interactions induced by dark matter particles which couple bilinearly to standard model particles through quantum fluctuations were termed quantum dark forces by Brax, Fichet, and Pignol, who constrained the mass and couplings of dark scalar, dark vector, and dark fermion bilinear exchange between nucleons. We extend their results to derive new constraints on quantum dark forces from bilinear scalar exchange between neutrons and electrons. 

\end{abstract}


\pacs{11.30.Er, 24.70.+s, 13.75.Cs}

\maketitle

\section {Introduction}

The existence of dark sector particles which induce relatively weak interactions among standard model particles is one of the many possibilities under active consideration to help address the dark matter and dark energy mysteries presented to us by astronomical and cosmological observations. Although one can imagine a very broad menagerie of such possibilities, several theoretical works over the last two decades have identified and classified specific mechanisms and scenarios with particular features which can be sought in laboratory experiments. 

If we keep the usual constraints which follow from the combination of relativity and quantum mechanics, one can proceed to analyze many of these possibilities within general frameworks that possess common mathematical forms for the interactions that can be applied to many physical systems. We expect these new interactions to be classifiable in terms of the strength, range, and (possible) spin dependence of the interactions. Exchange bosons with very weak couplings to ordinary matter and small but finite masses can induce forces over a wide range of distance scales. New interactions in this regime are now studied and analyzed intensively. Since many of the systems available to probe interactions in this regime undergo nonrelativistic motion, it is convenient to classify them in terms of nonrelativistic potentials and their corresponding scattering amplitudes. This limiting regime has been and continues to be explored in great detail, as one can see in several reviews~\cite{Adelberger:2009zz, Antoniadis:2011zza, Heil:2013tpa, SALUMBIDES201465, Murata_2015, Safronova:2017xyt, Lemos:2021xpi, Jiang_2025, RevModPhys.97.025005} 


The 2018 works~\cite{PhysRevLett.120.131801, PhysRevD.97.115034} of Brax, Fichet, and Pignol considered a qualitatively different possibility: the exchange of light (sub-GeV) dark sector particles of mass $m$ which couple bilinearly to standard model particles rather than linearly. Such interactions generate dark sector analogues of the well-known Casimir-Polder electromagnetic interaction between atoms~\cite{PhysRev.73.360} and the weak interaction from neutrino pair exchange~\cite{PhysRev.166.1638}. Since potentials from bilinear exchange possess a qualitatively different dependence on distance $r$ compared to single boson exchange, with Yukawa-type exponential falloff for $mr>>1$ and power-law falloff for $mr<<1$, the constraints one gets from the analysis of experiments are quite different from those which assume single boson exchange.  

The authors also showed that these interactions could be analyzed in a general way using an effective field theory approach valid for energy scales below $\Lambda_{QCD}$ and applied to operators of the form $O_{nuc}O_{DS}$, where $O_{DS}$ is a dark sector field bilinear and $O_{nuc}$ is a fermion bilinear of the form $\bar{N}\Gamma N$ with $N$ a fermion field and $\Gamma$ an arbitrary Lorentz structure. Reference~\cite{PhysRevLett.120.131801} analyzed constraints from spin-independent dark fermion and dark vector exchange, and reference~\cite{PhysRevD.97.115034} considered constraints from spin-independent dark scalar exchange. The details of the calculations for all cases were presented in an appendix in~\cite{PhysRevD.97.115034}. Later these results were extended to the case of spin-dependent quantum dark forces between nucleons~\cite{Costantino:2019ixl}.  

As suggested by the subscript on $O_{nuc}$, Fichet, Brax, and Pignol applied their work in~\cite{PhysRevD.97.115034} to the case of nucleon-nucleon interactions. They pointed out in section 2 of~\cite{PhysRevD.97.115034} that one could also apply their results to the case of electron-nucleon interactions, but they did not explicitly carry out this analysis.



In this paper we apply their work to constrain neutron-electron quantum dark forces from bilinear scalar exchange. We make use of a combination of data from neutron measurements at different scales to set constraints on neutron-electron quantum dark forces over a broad range of distance scales. As in the works~\cite{PhysRevLett.120.131801, PhysRevD.97.115034}, we can view this work as complementary to the analysis of exotic single boson exchange between nucleons. 

The rest of this paper is organized as follows. Section 2 reviews the calculation and results for the spin-independent quantum dark force potentials and scattering amplitudes required for our work. Section 3 reviews standard model neutron-electron interactions and measurements of the neutron-electron scattering amplitude and the neutron polarizability listed in the Particle Data Group averages. Section 4 presents the results. We conclude and suggest possibilities for additional work in Section 5. Documentation of derivations, normalization checks, and numerical robustness tests are collected in the Appendices. 


\section{Calculation of scalar quantum dark force potentials and scattering amplitudes}
\label{sec:qdf}


Following Ref.~\cite{PhysRevD.97.115034}, for a fermion spinor field \(f=n,e\) we consider the operators
\begin{equation}
\begin{aligned}
\Oa^{\,f}
&=
\frac{1}{\Lambda_{f,a}}\,
\bar f f\,\frac{\phi^2}{2},
\\[1mm]
\Ob^{\,f}
&=
\frac{1}{\Lambda_{f,b}^{\,2}}\,
\bar f\gamma^\mu f\,
\phi^\ast i \lrpartial_{\mu}\phi,
\\[1mm]
\Oc^{\,f}
&=
\frac{1}{\Lambda_{f,c}^{\,3}}\,
\bar f f\,
\frac{(\partial_\mu\phi)(\partial^\mu\phi)}{2}.
\end{aligned}
\label{eq:operators}
\end{equation}
where \(\phi\) is real for \(\Oa\) and \(\Oc\), and complex for
\(\Ob\). Each of the operators $O_{a}^{f}$, $O_{b}^{f}$, and $O_{c}^{f}$ is accompanied by a dimensionful coupling in an effective field theory treatment in terms of lepton and nucleon degrees of freedom coming from some more fundamental interaction at an energy scale $\Lambda$. For a neutron--electron loop, we separate out this common scale $\Lambda$ from the dimensionless neutron-scalar couplings $c_{n}^{i}$ and the electron-scalar couplings $c_{e}^{i}$ to get 

\begin{equation}
\Lambda_{ne,i}
=
\sqrt{\Lambda_{n,i}\Lambda_{e,i}},
\qquad
\alpha_i
=
\frac{s_i}{\Lambda_{ne,i}^{2k_i}}
=
\frac{c_n^{(i)}c_e^{(i)}}{\Lambda^{2k_i}},
\label{eq:alpha}
\end{equation}
where \((k_a,k_b,k_c)=(1,2,3)\) and we denote the relative sign of the neutron and electron Wilson coefficients by
\(s_i=\pm1\). We quote the magnitude as
\begin{equation}
\frac{1}{\Lambda_{ne,i}}
=
|\alpha_i|^{1/(2k_i)}.
\label{eq:Lne}
\end{equation}

The only difference compared to the nucleon-nucleon computation presented in~\cite{PhysRevD.97.115034} is that the loop for neutron-electron interactions contains one neutron and one electron vertex and is therefore proportional to the product \(c_nc_e\) rather than the positive combination \(c_n^2\). 


\subsection{Potentials}

The procedure for the calculation of the bubble diagram associated with the lowest-order process induced by these operators follows the standard Feynman rules using dimensional regularization as presented in quantum field theory texts. The loop basis, the dispersion relation, and the explicit calculation of the long-distance behaviors for all of these operators was derived and presented in Ref.~\cite{PhysRevD.97.115034}. We reproduce them in in Appendix~\ref{app:loops}. 

With the replacement
\begin{equation}
\frac{1}{\Lambda^{2k_i}}
\longrightarrow
\frac{s_i}{\Lambda_{ne}^{2k_i}},
\label{eq:replacement}
\end{equation}
the neutron--electron coordinate-space potentials have the same mathematical form as in Ref.~\cite{PhysRevD.97.115034}:
\begin{align}
V_a(r)
&=
-\frac{s_a}{32\pi^3\Lambda_{ne}^{2}}\,
\frac{m}{r^2}K_1(2mr),
\label{eq:Va}\\
V_b(r)
&=
\frac{s_b}{8\pi^3\Lambda_{ne}^{4}}\,
\frac{m^2}{r^3}K_2(2mr),
\label{eq:Vb}\\
V_c(r)
&=
-\frac{s_c}{32\pi^3\Lambda_{ne}^{6}r}
\Bigg[
\left(
\frac{30m^2}{r^4}+\frac{6m^4}{r^2}
\right)K_2(2mr)
\nonumber\\
&\hspace{2.5cm}
+
\left(
\frac{15m^3}{r^3}+\frac{m^5}{r}
\right)K_1(2mr)
\Bigg].
\label{eq:Vc}
\end{align}

At short distances \(mr\ll1\),
\begin{align}
V_a(r)&=-\frac{s_a}{64\pi^3\Lambda_{ne}^{2}r^3},
\nonumber\\
V_b(r)&=\frac{s_b}{16\pi^3\Lambda_{ne}^{4}r^5},
\nonumber\\
V_c(r)&=-\frac{15s_c}{32\pi^3\Lambda_{ne}^{6}r^7},
\label{eq:short}
\end{align}
while for \(mr\gg1\),
\begin{align}
V_a(r)
&\simeq
-\frac{s_a\sqrt m\,e^{-2mr}}
{64\pi^{5/2}\Lambda_{ne}^{2}r^{5/2}},
\nonumber\\
V_b(r)
&\simeq
\frac{s_bm^{3/2}e^{-2mr}}
{16\pi^{5/2}\Lambda_{ne}^{4}r^{7/2}},
\nonumber\\
V_c(r)
&\simeq
-\frac{s_cm^{9/2}e^{-2mr}}
{64\pi^{5/2}\Lambda_{ne}^{6}r^{5/2}} .
\label{eq:long}
\end{align}

In the $mr<<1$ limit the potentials from $O_{a}$, $O_{b}$, and $O_{c}$ are proportional to $1/r^{3}$, $1/r^{5}$, and $1/r^{7}$, respectively, and in the $mr>>1$ limit they vary as $\exp{(-2mr)}/r^{5/2}$, $\exp{(-2mr)}/r^{7/2}$, and $\exp{(-2mr)}/r^{9/2}$, respectively. 

\subsection{Scattering amplitudes}



The qualitatively-different distance dependence of these potentials compared to the heavily-analyzed case of single boson exchange creates a different landscape for the relative sensitivity of different physical systems to quantum dark forces. This fact encouraged us to analyze data on neutron-electron interactions as the distance scales probed by this data fall between nuclear and atomic distances. 

In our case we will use the neutron-electron scattering amplitude data, so we need the scattering amplitudes for the potentials. Following~\cite{PhysRevLett.120.131801, PhysRevD.97.115034}, we can constrain the different neutron-electron exotic interaction terms by fitting the data for the contribution of the different terms to the neutron-electron scattering length, $l_{NP}(Q)=l_{NP}^{C}(Q=0) + l_{NP}^{NC}(Q)$, where $l_{NP}^{C}(Q=0)$ is the \lq\lq contact\rq\rq piece of the new physics contribution, a constant independent of $Q$, and $l_{NP}^{NC}(Q)$ is the \lq\lq non-contact\rq\rq piece of the new physics contribution, which possesses a nontrivial Q-dependence. Here we use the same notation as in~\cite{PhysRevD.97.115034} and present the relevant expressions using the standard first-Born neutron-scattering normalization derived in Appendix~\ref{app:norm}. The full momentum-dependent loop expressions for the non-contact pieces of relevance in our analysis are
\begin{equation}
l_{{\rm NP},a}(Q)
=
\frac{s_am_n}{64\pi^3\Lambda_{ne}^{2}}\,f_0(Q),
\label{eq:la}
\end{equation}
\begin{equation}
l_{{\rm NP},b}(Q)
=
\frac{s_bm_n}{16\pi^3\Lambda_{ne}^{4}}
\left[
m^2f_0(Q)+Q^2f_1(Q)
\right],
\label{eq:lb}
\end{equation}
and
\begin{align}
l_{{\rm NP},c}(Q)
&=
\frac{s_cm_n}{64\pi^3\Lambda_{ne}^{6}}
\Bigg[
\left(3m^4-\frac{m^2Q^2}{2}\right)f_0(Q)
\nonumber\\
&\qquad+
\left(\frac{Q^4}{2}-12m^2Q^2\right)f_1(Q)
+10Q^4f_2(Q)
\Bigg].
\label{eq:lc}
\end{align}


with the non-contact kernels
\begin{align}
K_a^{NC}(Q,m)&=2L_s(Q,m),
\label{eq:Ka}\\
K_b^{NC}(Q,m)&=\frac{4m^2+Q^2}{3}L_s(Q,m),
\label{eq:Kb}\\
K_c^{NC}(Q,m)&=
\frac{5Q^4-40m^2Q^2+108m^4}{6}L_s(Q,m),
\label{eq:Kc}
\end{align}
with
\begin{equation}
L_s(Q,m)
=
\frac{1}{2}
\sqrt{1+\frac{4m^2}{Q^2}}
\ln\!\left[
\frac{\sqrt{1+4m^2/Q^2}+1}
{\sqrt{1+4m^2/Q^2}-1}
\right]-1 .
\label{eq:Ls}
\end{equation}
Thus
\begin{align}
C_a&=\frac{m_n}{64\pi^3}K_a^{NC},
\nonumber\\
C_b&=\frac{m_n}{16\pi^3}K_b^{NC},
\nonumber\\
C_c&=\frac{m_n}{64\pi^3}K_c^{NC},
\label{eq:C}
\end{align}
and
\begin{equation}
l_{{\rm NP},i}^{NC}(Q)
=
\alpha_i C_i(Q,m).
\label{eq:lNC}
\end{equation}
The complete \(f_0,f_1,f_2\) expressions and their spacelike continuation
are collected in Appendix~\ref{app:loops}.

\section {Review of standard model neutron-electron interactions}

To prepare for the discussion of the limits we will set later, we present a brief review of standard model neutron-electron interactions. In the slow neutron regime with $kR\ll1$ where $k$ is the neutron wave vector and $R$ is the range of the neutron-nucleus strong interaction for unpolarized neutrons incident upon unpolarized atoms with energies far from neutron-nucleus resonances, the neutron-atom scattering amplitude $b_\mathrm{atom}(q)$ as a function of the momentum transfer $q$ can be expressed as~\cite{Searsbook}


\begin{equation}
\label{batom}
b_\mathrm{atom}(q,E) = b(q,E)-b_{ne}Z[1-f(q)] +b_{pol}(q)\,.
\end{equation}

Other effects from standard model interactions are small compared to these terms. The first term $b(q,E)$ describes the low-energy coherent neutron scattering from the nucleus of the atom from the neutron-nucleus strong interaction, and the second and third terms come from the neutron charge radius and the neutron polarizability, respectively. The second term describes the interaction between the internal radial charge density of the neutron and the electric field of the atom. It is proportional to the bound neutron electron scattering length $b_{ne}$. $b_{ne}$ is directly proportional to the neutron mean-square charge radius $r_{n}^{2}$ through the relation


\begin{equation}
\rn
=
\frac{3m_ea_0}{m_n}\,\bne .
\label{eq:r2}
\end{equation}

Note that the neutron-electron scattering length used in neutron optics is a bound scattering length normalized with the neutron-atom reduced mass.

The $q$-dependence of $b_\mathrm{atom}(q,E)$ for slow neutron scattering comes from the atomic form factor $f(q)$ of the electron distribution around a nucleus of charge $Z$, which is accurately measured by x-ray scattering. The third term $b_\mathrm{pol}(q)$ is proportional to the very small but nonzero electric polarizability of the neutron and comes from the neutron electric dipole moment induced by the very large electric field near and inside the nucleus. It is directly related to the neutron electric polarizability $\alpha_{n}$. 

Although both $b_{ne}Z[1-f(q)]$ and $b_\mathrm{pol}(q)$ depend on momentum transfer $q$, the distance scales over which they vary significantly are set by the atomic and nuclear charge form factors, respectively. For $kR<<1$, $b_{pot}$ is independent of the incident neutron energy up to corrections of order $(kR)^{2}$. The slow neutron energy regime corresponds to $kR \approx 10^{-4}$ where $R$ is the range of the neutron-nucleus interaction. Therefore, as long as possible contributions from neutron-nucleus resonances are small, slow neutron measurements can determine $b_{ne}$ and $b_\mathrm{pol}(q)$.

$r_{n}^{2}$ has been derived from neutron scattering measurements of $b_{ne}$. We describe below the results listed in the Particle Data Group (PDG)~\cite{ParticleDataGroup:2026aaa} average. Measurements of the $q$-dependent angular distribution of thermal neutron scattering (25 meV) from noble-element gases neon, argon, krypton, and xenon~\cite{PhysRev.148.1303, PhysRevD.8.1305} determine $r_{n}^{2}=-0.115 \pm 0.003$ fm$^{2}$. 
Koester et al.~\cite{PhysRevC.51.3363} reported $r_{n}^{2}=-0.114 \pm 0.003$ fm$^{2}$ in scattering from lead and bismuth in the thermal to 2000 eV neutron energy range. Kopecky et al.~\cite{PhysRevC.56.2229} reported $r_{n}^{2}=-0.115 \pm 0.002 [stat.] \pm 0.003 [sys] $ fm$^{2}$ from neutron scattering in lead and $r_{n}^{2}=-0.124 \pm 0.003 [stat.] \pm 0.005 [sys] $ fm$^{2}$ from neutron scattering in bismuth in the thermal to 800 eV neutron energy range. 
With these measurements the current PDG average gives $r_{n}^{2}=-0.1155 \pm 0.0017$ fm$^{2}$.

Additional data on both $r_{n}^{2}$ and $\alpha_{n}$ from higher energies is in agreement with the neutron results. Although we do not use this data for our derived constraints on neutron-electron quantum dark forces, we mention this work for completeness as it sheds light on the neutron-electron results from very different experiments. 
For the neutron mean square charge radius, Atac et al.~\cite{Atac:2020hdq, Atac:2021wqj} report $r_{n}^{2}=-0.110 \pm 0.008$ fm$^{2}$ based on the extraction of the neutron electric form factor $G^{n}_{E}(Q^{2})$, at low four momentum
transfer squared ($Q^{2}$), where $r_{n}^{2}=-6\frac{dG^{n}_{E}(Q^{2})}{dQ^{2}}$ in the $Q^{2} \to 0$ limit. This extraction exploits a relation between the ratio of the quadrupole to the magnetic dipole transition form factors of the proton in $N \to \Delta$ transitions and the ratio $G_{n}
(E)/G_{n}(M)$ of the neutron electric and magnetic elastic form factors. Filin et al.~\cite{PhysRevLett.124.082501} report $r_{n}^{2}=-0.106 +0.007 / -0.005$ fm$^{2}$ based on the determination of the deuteron structure radius from chiral effective field theory, which uses high-precision atomic spectroscopy data to measure the difference in the deuteron and proton charge radii, which is dominated by the neutron charge radius. Both of these evaluations of $r_{n}^{2}$ are in agreement with the neutron data within errors. 

Two distinct determinations of the neutron electric polarizability employ Compton scattering from the deuteron. Myers et al.~\cite{COMPTONMAX-lab:2014cve} report $\alpha_{n}= [11.55 \pm 1.25 (stat.) \pm 0.8 (sys.)] \times 10^{-4}$ fm${^3}$ from an analysis of elastic Compton scattering from the deuteron, and Kossert et al~\cite{Kossert:2002ws} report $\alpha_{n}= [12.5 \pm 1.8 (stat.) +1.6/-1.3 (sys.)] \times 10^{-4}$ fm$^{3}$ from inelastic Compton scattering from the deuteron. Both of these determinations are in agreement with the neutron scattering measurement by Schmiedmayer et al~\cite{PhysRevLett.61.1065, PhysRevLett.61.2509} which reported $\alpha_{n}= [12.0 \pm 1.5 (stat.) \pm 2.0 (sys.)] \times 10^{-4}$ fm$^{3}$ from an analysis of a total cross section measurement in n-$^{208}$Pb scattering over the neutron energy range  1 eV-40 keV. All are included in the latest PDG average along with the less precise deuteron Compton scattering measurement from Rose et al~\cite{Rose1990621, Rose1990460} of  $\alpha_{n}= [10.7 (+3.3/-10.7) ] \times 10^{-4}$ fm$^{3}$. The agreement among these independent determinations of $\alpha_{n}$ retroactively confirms certain systematic error estimates made in the analysis of the $b_{ne}$ neutron measurements. 
\section {Constraints on neutron-electron quantum dark forces}

The new neutron-electron interaction enters an atomic observable with the electron form factor \(Zf(Q)\), while the corresponding electromagnetic charge-radius term is proportional to \(Z[1-f(Q)]\) where $f(Q) \to 1$ as $Q \to 0$ from the electrical neutrality of the atom. A nonzero nucleon coupling also generates a coherent nucleon-sector companion interaction, which must be accounted for when the neutron-electron interaction is constrained. Therefore the mixed neutron-electron force contributes
\begin{equation}
\delta b_{\rm atom}^{ne}(Q)
=
Zf(Q)\,\alpha_iC_i(Q,m).
\label{eq:neShape}
\end{equation}
and the same nucleon coupling also gives
\begin{equation}
\delta b_{\rm atom}^{NN}(Q)
=
A\,\beta_iC_i(Q,m),
\qquad
\beta_i\ge0.
\label{eq:NNShape}
\end{equation}



We performed a least-squares analysis of the data. We modeled the data using
\begin{equation}
\bne^{(r)}
=
b_0+\alpha_iR_r(m)+\beta_iS_r(m),
\label{eq:model}
\end{equation}

in terms of the response functions derived in the Appendices,
\begin{align}
R_r(m)
&=
\frac{\sum_jw_{rj}x_{rj}y^{ne}_{rj}}
{\sum_jw_{rj}x_{rj}^2},
\nonumber\\
S_r(m)
&=
\frac{\sum_jw_{rj}x_{rj}y^{NN}_{rj}}
{\sum_jw_{rj}x_{rj}^2}.
\label{eq:responses}
\end{align}

after defining  
\begin{equation}
x_{rj}=-Z_{rj}[1-f(Q_{rj})],
\end{equation}
\begin{equation}
y^{ne}_{rj}=Z_{rj}f(Q_{rj})C_i(Q_{rj},m),
\qquad
y^{NN}_{rj}=A_{rj}C_i(Q_{rj},m).
\end{equation}

for node \(j\) of determination \(r\). We minimized
\begin{equation}
\chi^2
=
\sum_r
\frac{
[\bne^{(r)}-b_0-\alpha_iR_r-\beta_iS_r]^2
}{\sigma_r^2},
\label{eq:chi2}
\end{equation}
over \(b_0\) and \(\beta_i\ge0\). Since \(\alpha_i\) can
have either sign, the 95\% confidence level interval is defined by
\begin{equation}
\chi^2(\alpha_i)-\chi^2(\hat\alpha_i)=3.84.
\end{equation}
For our limits we quote the magnitude
\begin{equation}
\frac{1}{\Lambda_{ne}}
=
\max(|\alpha_-|,|\alpha_+|)^{1/(2k_i)}.
\label{eq:limit}
\end{equation}

\begin{figure}[!t]
\centering
\includegraphics[width=\columnwidth]{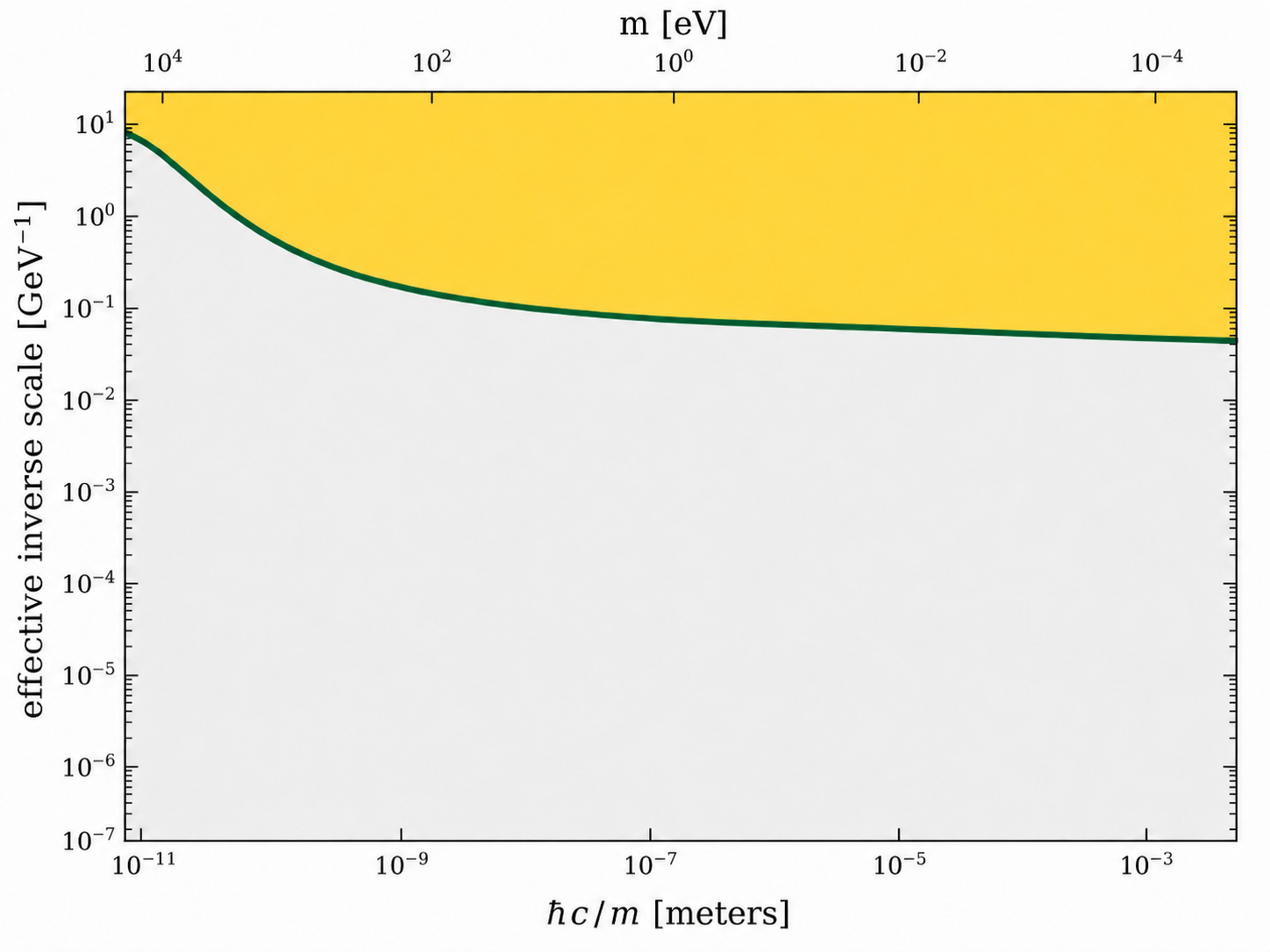}
\caption{Constraints on the effective inverse scale and mediator mass for
the operator \(\Oa\).}

\label{fig:Oa}
\end{figure}

\begin{figure}[!t]
\centering
\includegraphics[width=\columnwidth]{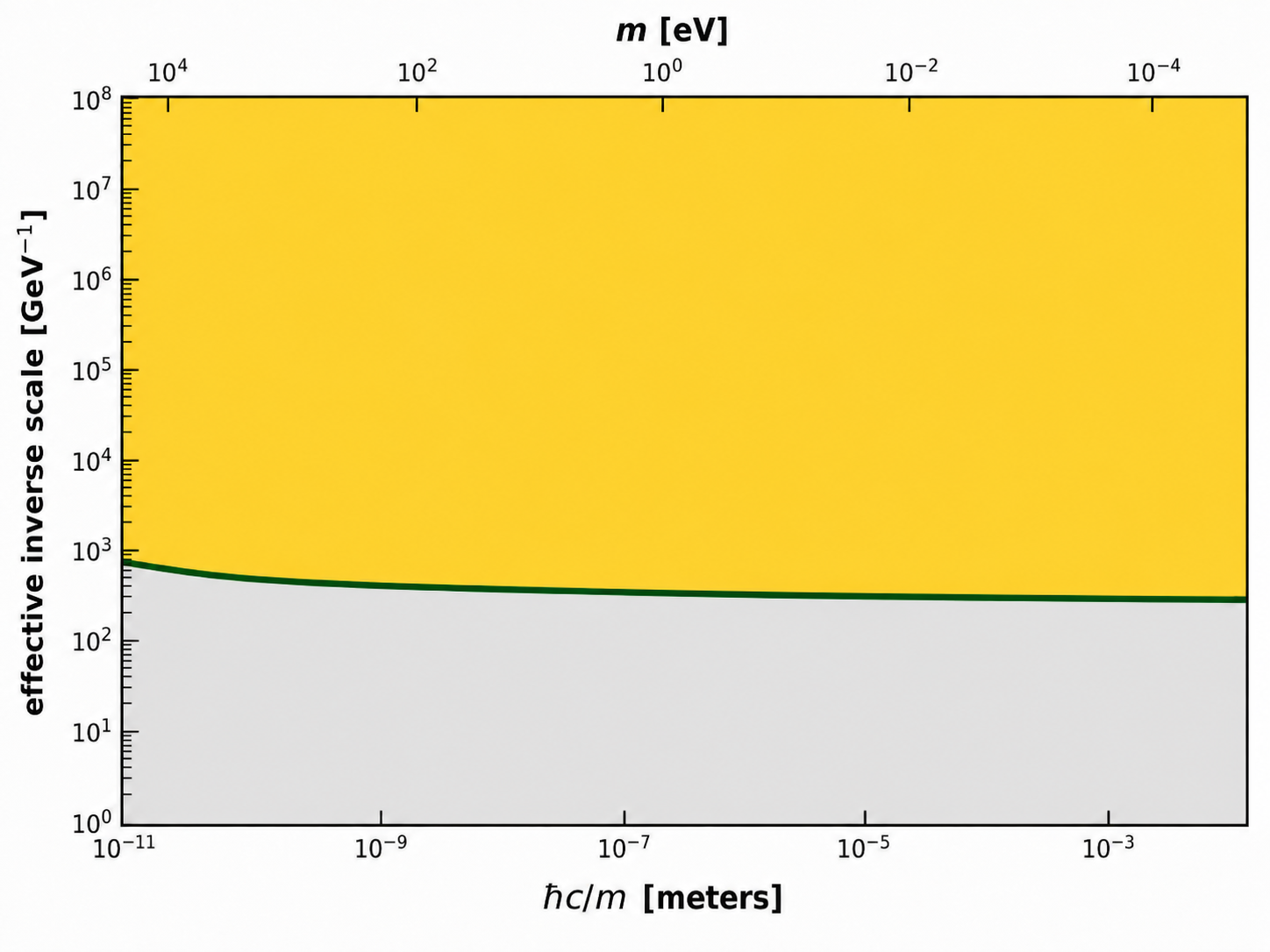}
\caption{Same as Fig.~\ref{fig:Oa}, but for the operator \(\Ob\).}
\label{fig:Ob}
\end{figure}

\begin{figure}[!t]
\centering
\includegraphics[width=\columnwidth]{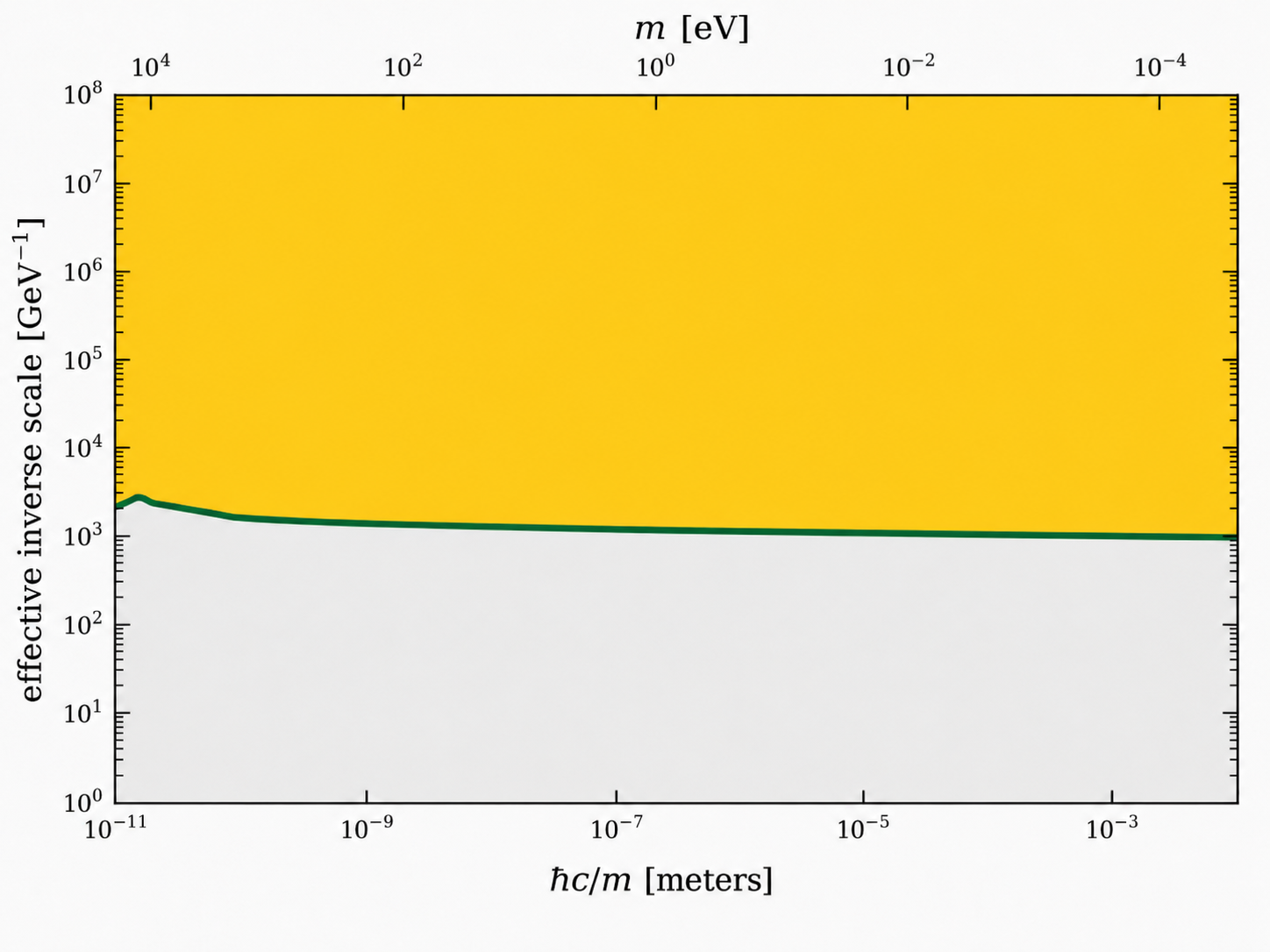}
\caption{Same as Fig.~\ref{fig:Oa}, but for the operator \(\Oc\).}
\label{fig:Oc}
\end{figure}

The limits are presented in Table~\ref{tab:limits} No deviation exceeds \(1.5\sigma\). The associated exclusion plots are shown in Figs.~\ref{fig:Oa}--\ref{fig:Oc}. 

\begin{table}[t]
\caption{95\% C.L. limits on \(1/\Lambda_{ne}\) in
GeV\(^{-1}\).}
\label{tab:limits}
\begin{ruledtabular}
\begin{tabular}{lccc}
\(m\) & \(\Oa\) & \(\Ob\) & \(\Oc\)\\
\hline
1 eV   & 0.478 & \(2.05\times10^2\) & \(1.14\times10^3\)\\
10 keV & 5.96  & \(4.38\times10^2\) & \(1.99\times10^3\)\\
\end{tabular}
\end{ruledtabular}
\end{table}

The higher-dimensional \(\Ob\) and \(\Oc\) operators are less constrained at the
few-keV momentum transfers relevant to the present neutron measurements because their non-contact pieces contain additional powers of momentum.

\section{Conclusion}



We have presented new limits on the mixed coupling product \(c_nc_e\) from quantum dark forces between neutrons and electrons based on an application of the work of Brax, Fichet, and Pignol to results on the neutron-electron scattering amplitude measurements included in the PDG average. 

With further theoretical work beyond the scope of this paper to calculate the sensitivity of the Compton scattering, transition form factor ratios, and proton and deuteron charge radii observables to neutron-electron quantum dark forces, it could be possible to extend our constraints to shorter distance scales. 

Many of the constraints presented in~\cite{PhysRevD.97.115034} on spin-independent bilinear dark scalar exchange between nucleons could be extended to bilinear electron-nucleon dark scalar exchange as well, but additional assumptions would need to be employed to isolate the electron-neutron component. 

As noted in~\cite{PhysRevLett.120.131801} quantum dark forces between electrons and nucleons could be studied by analyzing King plots from isotope shift spectroscopy~\cite{PhysRevD.96.093001, PhysRevLett.120.091801, PhysRevResearch.2.043444}. The experimental precision of isotope shift spectroscopy has rapidly improved over the last decade, and recent results ~\cite{PhysRevA.110.L030801, PhysRevLett.134.063002, PhysRevLett.134.233002} now require careful corrections to higher-order standard model interactions which can also lead to King plot nonlinearities. An analysis of these results within the quantum dark forces assumption could also yield improved constraints. Our results could also be of relevance for quantum dark force effects in the presence of background particles~\cite{Barbosa:2024pkl} and wake forces~\cite{PhysRevD.109.096036}.

\section{Acknowledgments}
\label{sec:ack}

This research was funded by the Science Committee of the Ministry of Science and Higher Education of the Republic of Kazakhstan (Grant No. AP23484023). W. M. Snow acknowledges support from US National Science Foundation grant PHY-2209481 and from the Indiana University Center for Spacetime Symmetries. We thank the authors of~\cite{PhysRevLett.120.131801, PhysRevD.97.115034} for sending the source files for some of the constraints in their paper.

\section{Loop functions, branch cut, and threshold behavior}
\label{app:loops}

The loop functions are written in the same basis as
Ref.~\cite{PhysRevD.97.115034},
\begin{equation}
f_n
=
\int_0^1dx\,[x(1-x)]^n
\ln\!\left(\frac{\Delta}{\Lambda^2}\right),
\qquad
\Delta=m^2-x(1-x)q^2.
\label{eq:fn}
\end{equation}
For elastic scattering \(q^2=-Q^2\), so
\begin{equation}
\Delta=m^2+x(1-x)Q^2>0
\end{equation}
and the loop functions are real. Their explicit spacelike forms are
\begin{align}
f_0(Q)
&=
2L_s+\ln\!\left(\frac{m^2}{\Lambda^2}\right),
\\
f_1(Q)
&=
\frac{Q^2-2m^2}{3Q^2}L_s
+\frac{1}{18}
+\frac{1}{6}\ln\!\left(\frac{m^2}{\Lambda^2}\right),
\\
f_2(Q)
&=
\frac{6m^4-2m^2Q^2+Q^4}{15Q^4}L_s
+\frac{13}{900}
-\frac{m^2}{30Q^2}
+\frac{1}{30}\ln\!\left(\frac{m^2}{\Lambda^2}\right).
\end{align}

The coordinate-space potential follows from the discontinuity across the
two-particle cut,
\begin{equation}
V(r)
=
\frac{i}{(2\pi)^2r}
\int_{2m}^{\infty}d\lambda\,
\lambda[\widetilde V]e^{-\lambda r}.
\end{equation}
The loop discontinuities are
\begin{align}
[f_0]
&=
i\pi\frac{2}{\lambda}\sqrt{\lambda^2-4m^2},
\\
[f_1]
&=
i\pi
\frac{2m^2+\lambda^2}{3\lambda^3}
\sqrt{\lambda^2-4m^2},
\\
[f_2]
&=
i\pi
\frac{6m^4+2m^2\lambda^2+\lambda^4}
{15\lambda^5}
\sqrt{\lambda^2-4m^2}.
\end{align}
The analytic \(Q^{2n}\ln(m/\Lambda)\) terms of the subtracted amplitude mix
with local derivative operators. The limits in the main text therefore
use only the nonanalytic branch-cut kernels.

For \(\Oc\), the least suppressed term in Eq.~\eqref{eq:Vc}, together
with \(K_\nu(z)\sim\sqrt{\pi/(2z)}e^{-z}\), yields
\[
V_c(r)\propto e^{-2mr}r^{-5/2},
\]
which agrees with the asymptotic form of the potential shown in the main text.

\section{Scattering-length normalization}
\label{app:norm}

The nonrelativistic potential is defined by
\begin{equation}
i\mathcal M
=
-i\widetilde V(q)(2m_n)(2m_e),
\end{equation}
so the external neutron and electron masses cancel in
\(\widetilde V\). For the Fourier convention
\[
V(r)
=
\int\frac{d^3q}{(2\pi)^3}
\widetilde V(\bm q)e^{i\bm q\cdot\bm r},
\]
the first-Born amplitude is
\begin{equation}
f(\bm q)
=
-\frac{\mu}{2\pi}\widetilde V(\bm q)
\qquad(\hbar=c=1).
\label{eq:born}
\end{equation}
For neutron scattering from a heavy atom,
\(\mu\simeq m_n\), which gives
\begin{equation}
\bne
=
\frac{m_n}{3m_ea_0}\rn
\end{equation}
and reproduces the bound neutron--electron scattering-length convention used in neutron optics.

Reference~\cite{PhysRevD.97.115034} identifies its neutron-scattering contribution
as \(l_{\rm NP}=2m_N\widetilde V\). For \(\mu\simeq m_N\), we get a factor of
\begin{equation}
\frac{2m_N}{m_N/(2\pi)}=4\pi.
\end{equation}
which is applied to the constraints constructed from neutron scattering lengths.

\section{Projection onto the measured scattering length}
\label{app:projection}

After adding to the standard model expression for the neutron-electron scatering length of the form
\begin{equation}
d_j=b_0+\bne x_j
\end{equation}
an additional small contribution \(\epsilon y_j\), we minimize the
weighted least-squares function 
\begin{equation}
\delta\bne
=
\epsilon\,
\frac{\sum_jw_jx_jy_j}
{\sum_jw_jx_j^2}.
\label{eq:proofProjection}
\end{equation}
For the mixed neutron-electron interaction
\(x_j=-Z_j[1-f(Q_j)]\) and
\(y_j=Z_jf(Q_j)C_i(Q_j,m)\) to give the  \(R_r\) expression.  An acceptance average of the pointwise ratio \(f/(1-f)\) would not be the appropriate response of the fitted coefficient.

Under the universal proton/neutron benchmark, the companion interaction is coherent over the nucleus and has the shape
\(A_j\beta_iC_i\), which gives the second response \(S_r\).
If proton and neutron Wilson coefficients were treated independently, the factor \(A\) would have to be be replaced by separate proton and neutron density contributions.

\section{Experimental reconstruction and robustness}
\label{app:numerics}

The scientific papers describing the \(\bne\) results provide aggregate determinations rather than reusable event-level likelihoods. In our response model we use a 25-meV Maxwellian distribution of neutron energies with equal Ne/Ar/Kr/Xe weights for the Krohn--Ringo work, a log-uniform 5-meV--2-keV neutron energy spectrum with equal
Pb/Bi weights for the Koester work, a log-uniform 0.08--800-eV neutron energy spectrum for the Kopecky Pb work, and representative nodes \(0.1,0.3,1,4\) eV for the Kopecky Bi work. The analytic atomic form factor is used through \(Q=25~\text{\AA}^{-1}\) and is not extrapolated beyond that range. The angular integrations use 288-point Gauss--Legendre quadrature.


\bibliographystyle{apsrev4-2}
\bibliography{ref}

@article{Adelberger:2009zz,
    author = "Adelberger, E. G. and Gundlach, J. H. and Heckel, B. R. and Hoedl, S. and Schlamminger, S.",
    title = "{Torsion balance experiments: A low-energy frontier of particle physics}",
    doi = "10.1016/j.ppnp.2008.08.002",
    journal = "Prog. Part. Nucl. Phys.",
    volume = "62",
    pages = "102--134",
    year = "2009"
}

@article{Antoniadis:2011zza,
    author = "Antoniadis, I. and others",
    title = "{Short-range fundamental forces}",
    doi = "10.1016/j.crhy.2011.05.004",
    journal = "Comptes Rendus Physique",
    volume = "12",
    pages = "755--778",
    year = "2011"
}

@article{Heil:2013tpa,
    author = "Heil, Werner and others",
    title = "{Spin clocks: Probing fundamental symmetries in nature}",
    doi = "10.1002/andp.201300048",
    journal = "Annalen Phys.",
    volume = "525",
    number = "8-9",
    pages = "539--549",
    year = "2013"
}

@article{SALUMBIDES201465,
title = {Bounds on fifth forces at the sub-Angstrom length scale},
journal = {Journal of Molecular Spectroscopy},
volume = {300},
pages = {65-69},
year = {2014},
note = {Spectroscopic Tests of Fundamental Physics},
issn = {0022-2852},
doi = {https://doi.org/10.1016/j.jms.2014.04.003},
url = {https://www.sciencedirect.com/science/article/pii/S0022285214000885},
author = {E.J. Salumbides and W. Ubachs and V.I. Korobov}
}

@article{Murata_2015,
doi = {10.1088/0264-9381/32/3/033001},
url = {https://doi.org/10.1088/0264-9381/32/3/033001},
year = {2015},
month = {jan},
publisher = {IOP Publishing},
volume = {32},
number = {3},
pages = {033001},
author = {Murata, Jiro and Tanaka, Saki},
title = {A review of short-range gravity experiments in the LHC era},
journal = {Classical and Quantum Gravity}
}

@article{Safronova:2017xyt,
    author = "Safronova, M. S. and Budker, D. and DeMille, D. and Kimball, Derek F. Jackson and Derevianko, A. and Clark, C. W.",
    title = "{Search for New Physics with Atoms and Molecules}",
    eprint = "1710.01833",
    archivePrefix = "arXiv",
    primaryClass = "physics.atom-ph",
    doi = "10.1103/RevModPhys.90.025008",
    journal = "Rev. Mod. Phys.",
    volume = "90",
    number = "2",
    pages = "025008",
    year = "2018"
}

@article{Lemos:2021xpi,
    author = "Lemos, A. S.",
    title = "{Submillimeter constraints for non-Newtonian gravity from spectroscopy}",
    eprint = "2105.14632",
    archivePrefix = "arXiv",
    primaryClass = "gr-qc",
    doi = "10.1209/0295-5075/135/11001",
    journal = "EPL",
    volume = "135",
    number = "1",
    pages = "11001",
    year = "2021"
}

@article{Jiang_2025,
doi = {10.1088/1361-6633/ad99e6},
url = {https://doi.org/10.1088/1361-6633/ad99e6},
year = {2024},
month = {dec},
publisher = {IOP Publishing},
volume = {88},
number = {1},
pages = {016401},
author = {Jiang, Min and Su, Haowen and Chen, Yifan and Jiao, Man and Huang, Ying and Wang, Yuanhong and Rong, Xing and Peng, Xinhua and Du, Jiangfeng},
title = {Searches for exotic spin-dependent interactions with spin sensors},
journal = {Reports on Progress in Physics}
}

@article{RevModPhys.97.025005,
  title = {Spin-dependent exotic interactions},
  author = {Cong, Lei and Ji, Wei and Fadeev, Pavel and Ficek, Filip and Jiang, Min and Flambaum, Victor V. and Guan, Haosen and Jackson Kimball, Derek F. and Kozlov, Mikhail G. and Stadnik, Yevgeny V. and Budker, Dmitry},
  journal = {Rev. Mod. Phys.},
  volume = {97},
  issue = {2},
  pages = {025005},
  numpages = {86},
  year = {2025},
  month = {Jun},
  publisher = {American Physical Society},
  doi = {10.1103/RevModPhys.97.025005},
  url = {https://link.aps.org/doi/10.1103/RevModPhys.97.025005}
}

@article{Costantino:2019ixl,
    author = "Costantino, Alexandria and Fichet, Sylvain and Tanedo, Philip",
    title = "{Exotic Spin-Dependent Forces from a Hidden Sector}",
    eprint = "1910.02972",
    archivePrefix = "arXiv",
    primaryClass = "hep-ph",
    reportNumber = "UCR-TR-2019-FLIP-NCC-1709",
    doi = "10.1007/JHEP03(2020)148",
    journal = "JHEP",
    volume = "03",
    pages = "148",
    year = "2020"
}

@article{PhysRevLett.120.131801,
  title = {Quantum Forces from Dark Matter and Where to Find Them},
  author = {Fichet, Sylvain},
  journal = {Phys. Rev. Lett.},
  volume = {120},
  issue = {13},
  pages = {131801},
  numpages = {6},
  year = {2018},
  month = {Mar},
  publisher = {American Physical Society},
  doi = {10.1103/PhysRevLett.120.131801},
  url = {https://link.aps.org/doi/10.1103/PhysRevLett.120.131801}
}

@article{PhysRevD.97.115034,
  title = {Bounding quantum dark forces},
  author = {Brax, Philippe and Fichet, Sylvain and Pignol, Guillaume},
  journal = {Phys. Rev. D},
  volume = {97},
  issue = {11},
  pages = {115034},
  numpages = {17},
  year = {2018},
  month = {Jun},
  publisher = {American Physical Society},
  doi = {10.1103/PhysRevD.97.115034},
  url = {https://link.aps.org/doi/10.1103/PhysRevD.97.115034}
}

@article{PhysRev.73.360,
  title = {The Influence of Retardation on the London-van der Waals Forces},
  author = {Casimir, H. B. G. and Polder, D.},
  journal = {Phys. Rev.},
  volume = {73},
  issue = {4},
  pages = {360--372},
  numpages = {0},
  year = {1948},
  month = {Feb},
  publisher = {American Physical Society},
  doi = {10.1103/PhysRev.73.360},
  url = {https://link.aps.org/doi/10.1103/PhysRev.73.360}
}

@article{PhysRev.166.1638,
  title = {Long-Range Forces from Neutrino-Pair Exchange},
  author = {Feinberg, G. and Sucher, J.},
  journal = {Phys. Rev.},
  volume = {166},
  issue = {5},
  pages = {1638--1644},
  numpages = {0},
  year = {1968},
  month = {Feb},
  publisher = {American Physical Society},
  doi = {10.1103/PhysRev.166.1638},
  url = {https://link.aps.org/doi/10.1103/PhysRev.166.1638}
}

@book{Searsbook,
  author    = {V. F. Sears},
  title     = {Neutron Optics: An Introduction to the Theory of Neutron Optical Phenomena and Their Applications},
  series    = {Oxford Series on Neutron Scattering in Condensed Matter},
  publisher = {Oxford University Press},
  year      = {1989}
}

@article{ParticleDataGroup:2026aaa,
    author = "Takahashi, F. and others",
    collaboration = "Particle Data Group",
    title = "{Review of Particle Physics}",
    doi = "10.1142/S0217751X26300115",
    journal = "Int. J. Mod. Phys. A",
    volume = "41",
    pages = "2630011",
    year = "2026"
}

@article{PhysRev.148.1303,
  title = {Measurement of the Electron-Neutron Interaction by the Asymmetrical Scattering of Thermal Neutrons by Noble Gases},
  author = {Krohn, V. E. and Ringo, G. R.},
  journal = {Phys. Rev.},
  volume = {148},
  issue = {4},
  pages = {1303--1311},
  numpages = {0},
  year = {1966},
  month = {Aug},
  publisher = {American Physical Society},
  doi = {10.1103/PhysRev.148.1303},
  url = {https://link.aps.org/doi/10.1103/PhysRev.148.1303}
}

@article{PhysRevD.8.1305,
  title = {Reconsideration of the Electron-Neutron Scattering Length As Measured by the Scattering of Thermal Neutrons by Noble Gases},
  author = {Krohn, V. E. and Ringo, G. R.},
  journal = {Phys. Rev. D},
  volume = {8},
  issue = {5},
  pages = {1305--1307},
  numpages = {0},
  year = {1973},
  month = {Sep},
  publisher = {American Physical Society},
  doi = {10.1103/PhysRevD.8.1305},
  url = {https://link.aps.org/doi/10.1103/PhysRevD.8.1305}
}

@article{PhysRevC.56.2229,
  title = {Neutron charge radius determined from the energy dependence of the neutron transmission of liquid ${}^{208}\mathrm{Pb}$ and ${}^{209}\mathrm{Bi}$},
  author = {Kopecky, S. and Harvey, J. A. and Hill, N. W. and Krenn, M. and Pernicka, M. and Riehs, P. and Steiner, S.},
  journal = {Phys. Rev. C},
  volume = {56},
  issue = {4},
  pages = {2229--2237},
  numpages = {0},
  year = {1997},
  month = {Oct},
  publisher = {American Physical Society},
  doi = {10.1103/PhysRevC.56.2229},
  url = {https://link.aps.org/doi/10.1103/PhysRevC.56.2229}
}

@article{PhysRevC.51.3363,
  title = {Neutron-electron scattering length and electric polarizability of the neutron derived from cross sections of bismuth and of lead and its isotopes},
  author = {Koester, L. and Waschkowski, W. and Mitsyna, L. V. and Samosvat, G. S. and Prokofjevs, P. and Tambergs, J.},
  journal = {Phys. Rev. C},
  volume = {51},
  issue = {6},
  pages = {3363--3371},
  numpages = {0},
  year = {1995},
  month = {Jun},
  publisher = {American Physical Society},
  doi = {10.1103/PhysRevC.51.3363},
  url = {https://link.aps.org/doi/10.1103/PhysRevC.51.3363}
}

@article{PhysRevLett.61.2509,
  title = {Measurement of the Electric Polarizability of the Neutron},
  author = {Schmiedmayer, J. and Rauch, H. and Riehs, P.},
  journal = {Phys. Rev. Lett.},
  volume = {61},
  issue = {21},
  pages = {2509},
  numpages = {0},
  year = {1988},
  month = {Nov},
  publisher = {American Physical Society},
  doi = {10.1103/PhysRevLett.61.2509},
  url = {https://link.aps.org/doi/10.1103/PhysRevLett.61.2509}
}

@article{PhysRevLett.61.1065,
  title = {Measurement of the Electric Polarizability of the Neutron},
  author = {Schmiedmayer, J. and Rauch, H. and Riehs, P.},
  journal = {Phys. Rev. Lett.},
  volume = {61},
  issue = {9},
  pages = {1065--1068},
  numpages = {0},
  year = {1988},
  month = {Aug},
  publisher = {American Physical Society},
  doi = {10.1103/PhysRevLett.61.1065},
  url = {https://link.aps.org/doi/10.1103/PhysRevLett.61.1065}
}

@article{Atac:2020hdq,
    author = "Atac, H. and Constantinou, M. and Meziani, Z. -E. and Paolone, M. and Sparveris, N.",
    title = "{Charge radii of the nucleon from its flavor dependent Dirac form factors}",
    eprint = "2009.04357",
    archivePrefix = "arXiv",
    primaryClass = "nucl-ex",
    doi = "10.1140/epja/s10050-021-00389-9",
    journal = "Eur. Phys. J. A",
    volume = "57",
    number = "2",
    pages = "65",
    year = "2021"
}

@article{Atac:2021wqj,
    author = "Atac, H. and Constantinou, M. and Meziani, Z. -E and Paolone, M. and Sparveris, N.",
    title = "{Measurement of the neutron charge radius and the role of its constituents}",
    eprint = "2103.10840",
    archivePrefix = "arXiv",
    primaryClass = "nucl-ex",
    doi = "10.1038/s41467-021-22028-z",
    journal = "Nature Commun.",
    volume = "12",
    number = "1",
    pages = "1759",
    year = "2021",
    note = "[Erratum: Nature Commun. 12, 3290 (2021)]"
}

@article{PhysRevLett.124.082501,
  title = {Extraction of the Neutron Charge Radius from a Precision Calculation of the Deuteron Structure Radius},
  author = {Filin, A. A. and Baru, V. and Epelbaum, E. and Krebs, H. and M\"oller, D. and Reinert, P.},
  journal = {Phys. Rev. Lett.},
  volume = {124},
  issue = {8},
  pages = {082501},
  numpages = {6},
  year = {2020},
  month = {Feb},
  publisher = {American Physical Society},
  doi = {10.1103/PhysRevLett.124.082501},
  url = {https://link.aps.org/doi/10.1103/PhysRevLett.124.082501}
}

@article{COMPTONMAX-lab:2014cve,
    author = "Myers, L. S. and others",
    collaboration = "COMPTON@MAX-lab",
    title = "{Measurement of Compton Scattering from the Deuteron and an Improved Extraction of the Neutron Electromagnetic Polarizabilities}",
    eprint = "1409.3705",
    archivePrefix = "arXiv",
    primaryClass = "nucl-ex",
    doi = "10.1103/PhysRevLett.113.262506",
    journal = "Phys. Rev. Lett.",
    volume = "113",
    number = "26",
    pages = "262506",
    year = "2014"
}

@article{Kossert:2002ws,
    author = "Kossert, K. and others",
    title = "{Quasifree Compton scattering and the polarizabilities of the neutron}",
    eprint = "nucl-ex/0210020",
    archivePrefix = "arXiv",
    reportNumber = "NUCL-EX-0210020",
    doi = "10.1140/epja/i2002-10093-9",
    journal = "Eur. Phys. J. A",
    volume = "16",
    pages = "259--273",
    year = "2003"
}

@ARTICLE{Rose1990621,
	author = {Rose, K.W. and Zurmahl, B. and Rullhusen, P. and Ludwig, M. and Baumann, A. and Schumacher, M. and Ahrens, J. and Zieger, A. and Christmann, D. and Ziegler, B.},
	title = {Quasi-free compton scattering by the neutron},
	year = {1990},
	journal = {Nuclear Physics, Section A},
	volume = {514},
	number = {4},
	pages = {621},
	doi = {10.1016/0375-9474(90)90014-D},
	url = {https://www.scopus.com/inward/record.uri?eid=2-s2.0-0000673002&doi=10.1016%2f0375-9474%2890%2990014-D&partnerID=40&md5=e9fd2dc572ae446626d26dc9ac8ce782},
	type = {Article},
	publication_stage = {Final},
	source = {Scopus},
}

@ARTICLE{Rose1990460,
	author = {Rose, K.W. and Zurmahl, B. and Rullhusen, P. and Ludwig, M. and Baumann, A. and Schumacher, M. and Ahrens, J. and Zieger, A. and Christmann, D. and Ziegler, B. and Schoch, B.},
	title = {Polarizability of the neutron},
	year = {1990},
	journal = {Physics Letters B},
	volume = {234},
	number = {4},
	pages = {460},
	doi = {10.1016/0370-2693(90)92039-L},
	url = {https://www.scopus.com/inward/record.uri?eid=2-s2.0-0001263320&doi=10.1016%2f0370-2693%2890%2992039-L&partnerID=40&md5=8b832522f3b9c65ffdf4b7dfada99c12},
	type = {Article},
	publication_stage = {Final},
	source = {Scopus},
}

@article{PhysRevD.96.093001,
  title = {Probing atomic Higgs-like forces at the precision frontier},
  author = {Delaunay, C\'edric and Ozeri, Roee and Perez, Gilad and Soreq, Yotam},
  journal = {Phys. Rev. D},
  volume = {96},
  issue = {9},
  pages = {093001},
  numpages = {7},
  year = {2017},
  month = {Nov},
  publisher = {American Physical Society},
  doi = {10.1103/PhysRevD.96.093001},
  url = {https://link.aps.org/doi/10.1103/PhysRevD.96.093001}
}

@article{PhysRevLett.120.091801,
  title = {Probing New Long-Range Interactions by Isotope Shift Spectroscopy},
  author = {Berengut, Julian C. and Budker, Dmitry and Delaunay, C\'edric and Flambaum, Victor V. and Frugiuele, Claudia and Fuchs, Elina and Grojean, Christophe and Harnik, Roni and Ozeri, Roee and Perez, Gilad and Soreq, Yotam},
  journal = {Phys. Rev. Lett.},
  volume = {120},
  issue = {9},
  pages = {091801},
  numpages = {7},
  year = {2018},
  month = {Feb},
  publisher = {American Physical Society},
  doi = {10.1103/PhysRevLett.120.091801},
  url = {https://link.aps.org/doi/10.1103/PhysRevLett.120.091801}
}

@article{PhysRevResearch.2.043444,
  title = {Generalized King linearity and new physics searches with isotope shifts},
  author = {Berengut, Julian C. and Delaunay, C\'edric and Geddes, Amy and Soreq, Yotam},
  journal = {Phys. Rev. Res.},
  volume = {2},
  issue = {4},
  pages = {043444},
  numpages = {11},
  year = {2020},
  month = {Dec},
  publisher = {American Physical Society},
  doi = {10.1103/PhysRevResearch.2.043444},
  url = {https://link.aps.org/doi/10.1103/PhysRevResearch.2.043444}
}

@article{PhysRevA.110.L030801,
  title = {Systematic-free limit on new light scalar bosons via isotope-shift spectroscopy in ${\mathrm{Ca}}^{+}$},
  author = {Chang, Timothy T. and Awazi, Bless Bah and Berengut, Julian C. and Fuchs, Elina and Doret, S. Charles},
  journal = {Phys. Rev. A},
  volume = {110},
  issue = {3},
  pages = {L030801},
  numpages = {6},
  year = {2024},
  month = {Sep},
  publisher = {American Physical Society},
  doi = {10.1103/PhysRevA.110.L030801},
  url = {https://link.aps.org/doi/10.1103/PhysRevA.110.L030801}
}

@article{PhysRevLett.134.063002,
  title = {Probing New Bosons and Nuclear Structure with Ytterbium Isotope Shifts},
  author = {Door, Menno and Yeh, Chih-Han and Heinz, Matthias and Kirk, Fiona and Lyu, Chunhai and Miyagi, Takayuki and Berengut, Julian C. and Biero\ifmmode \acute{n}\else \'{n}\fi{}, Jacek and Blaum, Klaus and Dreissen, Laura S. and Eliseev, Sergey and Filianin, Pavel and Filzinger, Melina and Fuchs, Elina and F\"urst, Henning A. and Gaigalas, Gediminas and Harman, Zolt\'an and Herkenhoff, Jost and Huntemann, Nils and Keitel, Christoph H. and Kromer, Kathrin and Lange, Daniel and Rischka, Alexander and Schweiger, Christoph and Schwenk, Achim and Shimizu, Noritaka and Mehlst\"aubler, Tanja E.},
  journal = {Phys. Rev. Lett.},
  volume = {134},
  issue = {6},
  pages = {063002},
  numpages = {7},
  year = {2025},
  month = {Feb},
  publisher = {American Physical Society},
  doi = {10.1103/PhysRevLett.134.063002},
  url = {https://link.aps.org/doi/10.1103/PhysRevLett.134.063002}
}

@article{PhysRevLett.134.233002,
  title = {Nonlinear Calcium King Plot Constrains New Bosons and Nuclear Properties},
  author = {Wilzewski, Alexander and Spie\ss{}, Lukas J. and Wehrheim, Malte and Chen, Shuying and King, Steven A. and Micke, Peter and Filzinger, Melina and Steinel, Martin R. and Huntemann, Nils and Benkler, Erik and Schmidt, Piet O. and Huber, Luca I. and Flannery, Jeremy and Matt, Roland and Stadler, Martin and Oswald, Robin and Schmid, Fabian and Kienzler, Daniel and Home, Jonathan and Craik, Diana P. L. Aude and Door, Menno and Eliseev, Sergey and Filianin, Pavel and Herkenhoff, Jost and Kromer, Kathrin and Blaum, Klaus and Yerokhin, Vladimir A. and Valuev, Igor A. and Oreshkina, Natalia S. and Lyu, Chunhai and Banerjee, Sreya and Keitel, Christoph H. and Harman, Zolt\'an and Berengut, Julian C. and Viatkina, Anna and Gilles, Jan and Surzhykov, Andrey and Rosner, Michael K. and Crespo L\'opez-Urrutia, Jos\'e R. and Richter, Jan and Mariotti, Agnese and Fuchs, Elina},
  journal = {Phys. Rev. Lett.},
  volume = {134},
  issue = {23},
  pages = {233002},
  numpages = {10},
  year = {2025},
  month = {Jun},
  publisher = {American Physical Society},
  doi = {10.1103/PhysRevLett.134.233002},
  url = {https://link.aps.org/doi/10.1103/PhysRevLett.134.233002}
}

@article{Barbosa:2024pkl,
    author = "Barbosa, Sergio and Fichet, Sylvain",
    title = "{Background-induced forces from dark relics}",
    eprint = "2403.13894",
    archivePrefix = "arXiv",
    primaryClass = "hep-ph",
    doi = "10.1007/JHEP01(2025)021",
    journal = "JHEP",
    volume = "01",
    pages = "021",
    year = "2025"
}

@article{PhysRevD.109.096036,
  title = {Wake forces in a background of quadratically coupled mediators},
  author = {Van Tilburg, Ken},
  journal = {Phys. Rev. D},
  volume = {109},
  issue = {9},
  pages = {096036},
  numpages = {25},
  year = {2024},
  month = {May},
  publisher = {American Physical Society},
  doi = {10.1103/PhysRevD.109.096036},
  url = {https://link.aps.org/doi/10.1103/PhysRevD.109.096036}
}

\end{document}